\documentclass{elsart}

\usepackage{graphicx}% Include figure files
\usepackage{dcolumn}% Align table columns on decimal point
\usepackage{bm}% bold math
\usepackage{natbib}
\usepackage{amssymb}
\begin{document}

\begin{frontmatter}
%\draft
\title{Langevin equation with Coulomb friction}
\author{Hisao Hayakawa}
%\footnote{The corresponding
%author:
%hisao@yuragi.jinkan.kyoto-u.ac.jp}}
%\email{hisao@yuragi.jinkan.kyoto-u.ac.jp}
%\affiliation
\address{Department of Physics, Yoshida-South Campus, Kyoto University,
Kyoto 606-8501, Japan}
%\affiliation

\begin{abstract}
We propose
 a Langevin model with Coulomb
 friction. Through the analysis of the corresponding Fokker-Planck
 equation, we have obtained the steady velocity distribution function 
under the influence of the external field. 

\end{abstract}

%\pacs{81.05.Rm,05.40.-a 05.20.Dd}

\end{frontmatter}

%\maketitle

\section{Introduction}

The Langevin equation is widely used to describe the motion of
stochastic particles.\citep{Kubo} 
In a typical situation a Brownian 
particle is driven by the random force originated
from collisions of molecules in the environment and
the friction force
proportional to the velocity of the particle.

The friction force is not always proportional to the velocity of
particles. A well-known counter example can be found in the dry friction of 
macroscopic materials, {\it so called } Coulomb friction in which the
dynamical friction is independent of the speed of a material  but
depends on the direction of the motion.\citep{perrson} 
In general,
the motion of
such the material obeying Coulomb friction is 
not affected by  the thermal fluctuations, 
because the material is too large to be fluctuated by the thermal force.
% and  the static
%friction is usually larger than the dynamical friction in the dry
%friction.
 However, if the material we consider is enough small like  a
 nanomaterial  or the thermal fluctuation is replaced by the mechanical 
 fluctuation force,
Coulomb friction force with fluctuating random force may be relevant. 
%to describe the macroscopic motion of the material. 
In a recent paper, in fact,
Kawarada and Hayakawa\citep{kawarada} suggest
 that Langevin equation associated
with Coulomb friction can be an effective equation in a vibrating
system of granular particles through their simulation of granular
particles. They demonstrate that (i) granular particles in a
quasi-two-dimensional container obey an exponential velocity
distribution function (VDF), (ii) the exponential VDF is produced by
Coulomb slip between particles and fixed scatters because their
simulation without Coulomb slip produces a Gaussian-like VDF, (iii)
Langevin equation with Coulomb friction produces an exponential VDF, and 
(iv) the relation between the diffusion coefficient $D_x$ and the granular
temperature $T$ in their simulation obeys $D_x\propto \sqrt{T}$ as is
expected from Langevin equation with Coulomb friction.   
Their result may encourage us to look for
relevancy of such Langevin equation in physical systems, and
investigate mathematical properties of the equation. 

In this paper, we investigate mathematical properties of Langevin
equation with Coulomb friction.
% and compare the result of the
%calculation with the simulations of physical systems. 
The organization of this paper is as follows. In the next section, we
introduce our model. In section 3, we obtain the steady VDF under the
influence of the steady external field. In section 4 we discuss our
result through the comparison of our result with the simulation of
driven granular particles in a quasi-two-dimensional box. In the final
section, we conclude our results.

\section{Langevin model}

Let us consider the motion of a particle whose velocity is ${\bf v}$
subjected to the dynamical Coulomb friction force $-\zeta m {\bf
v}/|{\bf v}| $ with the mass $m$. When we
assume that the motion is affected by the external force ${\bm F}_{ex}$ and
the uncorrelated fast agitation by the environment, 
the equation of motion is given by
\begin{equation}\label{eq1}
\frac{d{\bf v}}{dt}=-\zeta \frac{{\bf v}}{|{\bf v}|}+\frac{{\bm F}_{ex}}{m}+{\bm \xi}
\end{equation}
where 
${\bm \xi}$ is the random force whose
$\alpha$ component is
assumed to satisfy
\begin{equation}\label{eq2}
<{\xi}_{\alpha}(t)>=0, \quad 
<{\xi}_{\alpha}(t)\xi_{\beta}(t')>=2D\delta_{\alpha,\beta}\delta(t-t').
\end{equation}
Here $<\cdots>$ represents the average over the distribution of the
random variable ${\bm \xi}$ and the diffusion constant $D$ in the velocity
space satisfies the fluctuation-dissipation relation
$D=\zeta\sqrt{T/(3m)}$, where $T$ is the temperature defined from the
second moment of VDF as $T(t)\equiv \frac{1}{2}m\int d{\bf v}v^2P({\bf
v},t)$. Since we assume that
higher order correlations of ${\bm\xi}$ vanish,
Langevin equation (\ref{eq1}) can be converted into 
Fokker-Planck equation for the probability function $P({\bf v},t)$:
\begin{equation}\label{eq3}
\frac{\partial P({\bf v},t)}{\partial t}+
\frac{{\bm F}_{ex}}{m}\cdot \frac{\partial}{\partial {\bf v}}P({\bf v},t)
=
\frac{\partial}{\partial {\bf v}}\cdot\left(\zeta\frac{\bf v}{|{\bf v}|}
P({\bf v},t)+D\frac{\partial}{\partial {\bf v}}P({\bf v},t)\right).
\end{equation}
Here $P({\bf v},t)$ satisfies the normalization condition
\begin{equation}\label{eq-norm}
\int d{\bf v} P({\bf v},t)=1 .
\end{equation}
%If we replace ${\bf v}/|{\bf v}|$ by ${\bf v}$, eq.(\ref{eq3}) is
%reduced to the standard Fokker-Planck equation in which the equilibrium
%solution of $P({\bf v},t)$in the absence of ${\bm F}_{ex}$ is Maxwellian.  

Kawarada and Hayakawa\citep{kawarada} indicate that 
eq.(\ref{eq3}) without 
${\bm F}_{ex}$ has the steady solution obeying an exponential
VDF. Namely, the VDF approaches
\begin{equation}\label{steady-sol}
P({\bf v},t)\to \varphi_0(v)\equiv \frac{\eta^2}{2\pi}\exp[-\eta v]
\end{equation}
as time goes on, where $\eta=\zeta/D$.
Thus, our system has completely different properties from those in the
conventional Langevin model.

It is convenient to use dimensionless distribution function. For this
purpose, we may introduce
\begin{equation}\label{scale-1}
P({\bf v},t)=v_0(t)^{-d}\tilde P({\bf c},t), \quad {\bf c}={\bf v}/v_0(t)
\end{equation}
with the normalization 
\begin{equation}\label{scale-2}
\int d{\bf c}c^2 \tilde{P}_0({\bf c},t)=1
\end{equation}
where $d$ is the dimension, and 
$\tilde{P}_0$ is the scaled VDF without ${\bm F}_{ex}$. 
We adopt $\tilde P_0$ instead of $\tilde P$ in
eq.(\ref{scale-2}), because the determination of the scaled factor is
difficult if we use $\tilde P$. In fact, $T$ is the function of ${\bm F}_{ex}$ and
cannot be determined without the complete form of VDF.

\section{Steady state distribution function}

Let us consider the situation that a particle is moving under the
influence of ${\bm F}_{ex}=mg \hat z$ with the unit vector $\hat z$ parallel to
the direction of the external force 
and Coulomb friction force associated with the random force.  
We assume that the system is a two-dimensional one, because the particle 
is typically located on a substrate when Coulomb friction is important.
We are interested in the statistical steady state in the balance between the
friction and the external force.

Let $\theta$ be the
angle between the direction we consider and $\hat z$, 
the steady equation becomes
\begin{equation}\label{eq2-1} 
g[\cos\theta \frac{\partial P}{\partial v}+\frac{\sin^2\theta}{v}
\frac{\partial P}{\partial \cos\theta}]
=\zeta\{\frac{P}{v}+\frac{\partial P}{\partial v}\}+
D[\frac{\partial^2}{\partial v^2}+\frac{1}{v}\frac{\partial}{\partial v}
+\frac{1}{v^2}\frac{\partial^2}{\partial \theta^2}]P,
\end{equation}
where $v=|{\bf v}|$.

Now we assume the expansion
\begin{equation}\label{eq2-2}
P(v,\theta)=\sum_{n=0}^{\infty}P_n(v)\cos n\theta=P_0(v)+P_1(v)\cos\theta.
\end{equation}
Here the terms with $n\ge 2$ are irrelevant,
because the contribution of $n\ge 2$ is orthogonal to those of $n=0,1$,
and the effect of gravity appears in the term of $n=1$. Thus,
the normalization (\ref{eq-norm}) is reduced to
\begin{equation}\label{eq2-2a}
2\pi\int_0^{\infty} dv vP_0(v)=1,
\end{equation}
and
eq.(\ref{eq2-1}) becomes
\begin{eqnarray}\label{eq2-3}
g &[& \cos\theta P_0'+\cos^2\theta P_1'+\frac{\sin^2\theta}{v}P_1]
=\zeta[\frac{P_0}{v}+P_0'+(\frac{P_1}{v}+P_1')\cos\theta] \nonumber \\
& &+D[P_0''+\frac{P_0'}{v}+(P_1''+\frac{P_1'}{v})\cos\theta 
-\frac{P_1}{v^2}
\cos\theta],
\end{eqnarray}
where $P_n'$ and $P_n''$ represent $dP_n/dv$ and $d^2P_n/dv^2$, respectively.
From the integrations of eq.(\ref{eq2-3}) 
multiplied by $\cos n\theta$ with $n=0,1$ over $(0,2\pi)$, we obtain
\begin{eqnarray}\label{eq2-4a}
g(P_1'+\frac{P_1}{v})&=&\zeta(\frac{P_0}{v}+P_0')+D(P_0''+\frac{P_0'}{v}) ,
 \\
g P_0'&=& \zeta(\frac{P_1}{v}+P_1')+D(P_1''+\frac{P_1'}{v}-\frac{P_1}{v^2}).
\label{eq2-4b}
\end{eqnarray}
Equation (\ref{eq2-4a}) can be integrated easily as
\begin{equation}\label{eq2-5}
P_1=\frac{\zeta}{g}P_0+\frac{D}{g}P_0'
\end{equation}
for $g\ne 0$. We can check that $P_0\propto e^{-\eta v}$ and
$P_1=0$ are the solution of (\ref{eq2-4a}) and (\ref{eq2-4b}) which
correspond to the steady solution (\ref{steady-sol}) at $g=0$.
%Thus, the equilibrium distribution function is not the Gaussian but the
%exponential.

To solve eq.(\ref{eq2-4b})
we adopt the form
\begin{equation}\label{eq2-6}
P_0(v)=f(v)e^{-\eta v}
\end{equation}
and substitute this into (\ref{eq2-5}), eq.(\ref{eq2-5}) is reduced to
\begin{equation}\label{eq2-7}
P_1(v)=\frac{D}{g}f'(v)e^{-\eta v}.
\end{equation}
With the aid of eqs.(\ref{eq2-6}) and
(\ref{eq2-7}), eq.(\ref{eq2-4b}) can be rewritten as
\begin{equation}\label{eq2-8}
f'''+\left(\frac{1}{v}-\eta\right)f''-
\left(\frac{1}{v^2}+\epsilon \eta^2\right)f'+\epsilon \eta^3 f=0 ,
\end{equation}
where $\epsilon=(g/\zeta)^2$.

Since we do not know the general procedure to obtain the 
solution of the third order ordinary differential
equation (\ref{eq2-8}) and it may not be easy to  get the numerical solution
for eq.(\ref{eq2-8}) around the singular point $v=0$, 
we rewrite it as a set of the second order
differential equations under the assumption of small $\epsilon$.  For
this purpose, we expand
\begin{equation}\label{eq2-9} 
f(v)=\frac{\eta^2}{2\pi}[1+\epsilon f^{(1)}+\epsilon^2 f^{(2)}+\cdots ].
\end{equation}
Here the prefactor is determined by the normalization (\ref{eq2-2a}).
 Thus, we obtain the series of
equations:
\begin{eqnarray}\label{eq2-10a}
& &h_1''+(\frac{1}{v}-\eta)h_1'-\frac{1}{v^2}h_1 =-\eta^3
 \\
& &h_n''+(\frac{1}{v}-\eta)h_n'-\frac{1}{v^2}h_n =-\eta^3 f^{(n-1)}+\eta^2h_{n-1}\quad {\rm for{~}}n\ge 2
\label{eq2-10b}\end{eqnarray}
with $h_n(v)\equiv df^{(n)}(v)/dv$.
We also assume
\begin{equation}\label{eq2-11}
\int_0^{\infty} dv v f^{(n)}(v)e^{-\eta v}=0\quad {\rm for{~}}n\ge 1,
\end{equation}
which ensures that the normalization condition (\ref{eq2-2a})
 is satisfied for any $\epsilon$.

Equations (\ref{eq2-10a}) and (\ref{eq2-10b}) have a common homogeneous part 
supplemented by the inhomogeneous parts in the right hand sides.
Thus, at first, let us obtain the solution of the homogeneous
equation:
\begin{equation}\label{eq2-12}
w''+(\frac{1}{v}-\eta)w'-\frac{1}{v^2}w =0 .
\end{equation}
The solution of eq.(\ref{eq2-12}) 
can be obtained by the method of 
a series-expansion around the singular point $v=0$. The result is
\begin{eqnarray}\label{eq2-13a}
w_1(v)&=&2\sum_{n=0}^{\infty} \frac{\eta^n}{(n+2)!}v^{n+1}=
\frac{2}{\eta^2v}(e^{\eta v}-1-\eta v) \\
w_2(v)&=& \frac{\eta^2}{2}w_1(v)\ln v-\sum_{k=1}^{\infty}\frac{\eta^k}{k!}
\{\sum_{r=1}^k\frac{1}{r}\} v^{k-1} \nonumber \\
&=& \frac{\eta^2}{2}w_1(v)\ln v-\frac{e^{\eta v}}{v}\{\gamma+\ln (\eta v) 
-Ei(-\eta v)\}
\label{eq2-13b}
\end{eqnarray}
where $\gamma$ is Euler's constant
$\gamma=
\lim_{n\to\infty}(\sum_{r=1}^{\infty}\frac{1}{r}-\ln n)=0.57721\cdots$ and
\begin{equation}\label{eq2-14}
Ei(-x)=-\int_x^{\infty}dt\frac{e^{-t}}{t}
\end{equation}
is the exponential integral function. Here we use Frobenius' method and
Bessel's formula\citep{arfken}
\begin{equation}\label{eq2-15}
\sum_{k=1}^{\infty}\sum_{r=1}^k\frac{1}{r}\frac{x^k}{k!}
=e^x\{\gamma+\ln x-Ei(-x)\}.
\end{equation}

The general solutions of inhomogeneous equation (\ref{eq2-10a}) or (\ref{eq2-10b})
is given by the sum of a special solution of the inhomogeneous equation
and the linear combination of the homogeneous solutions 
(\ref{eq2-13a}) and (\ref{eq2-13b}). For $O(\epsilon)$ it is easy to confirm 
that the special solution $h_{1,S}$ is given by
\begin{equation}\label{eq2-16}
h_{1,S}(v)=\eta^2 v.
\end{equation}
On the other hand, the leading order contribution $e^{\eta v}/v$ should
be canceled in the linear combination of $w_1$ and $w_2$ to obtain the
finite result for $\int_0^{\infty}dv v f^{(1)}(v)e^{-\eta v}$. Thus, we
obtain the solution as
\begin{eqnarray}\label{eq2-17}
h_1(v)&=&c_0 \{w_2(v)+\frac{\eta^2}{2}(\gamma+\ln \eta) w_1(v)\}+\eta^2 v
\nonumber \\
&=& c_0\{\frac{Ei(-\eta v)}{v}e^{\eta v}-\frac{1+\eta v}{v}(\gamma+\ln(\eta v))\}+\eta^2 v
\end{eqnarray}
where $c_0$ is a constant. We should note that $h_1(v)$ approaches
$h_1(v)\to -c_0\eta$ in the limit of $v\to 0$.
 
From the integration we obtain the first order solution as
\begin{eqnarray}\label{eq2-18}
f^{(1)}(v)&=&c_0\{G^{31}_{23}\left(\eta v|\matrix{
0 & 1 & \cr
0 & 0 & 0 \cr }\right)-\eta v(\gamma-1)+(\gamma+\eta v)\ln \eta v
-\frac{1}{2}(\ln \eta v)^2\} \nonumber \\
& &+\frac{\eta^2}{2}v^2+c_1
\end{eqnarray}
where $c_1$ is a constant and
\begin{eqnarray}
G^{m,n}_{p,q}\left(z|\matrix{a_1,& \cdots & a_p \cr
b_1 & \cdots & b_q  }\right)
&\equiv& (2\pi i)^{-1}\int ds z^s
\prod_{j=1}^m\Gamma(b_j-s)\prod_{j=1}^n\Gamma(1-a_j+s) \nonumber \\
& &\times 
\left(\prod_{j=m+1}^q\Gamma(1-b_j+s)\prod_{j=n+1}^p\Gamma(a_j-s)\right)^{-1}
\end{eqnarray}
is Meijer's G function\citep{meijer} which satisfies
\begin{equation}
\frac{d}{dz}G^{31}_{23}\left(z|\matrix{
0 & 1 & \cr
0 & 0 & 0 \cr }\right)
=\frac{e^z}{z}Ei(-z) .
\end{equation}
From the normalization condition (\ref{eq2-11}) 
constants $c_0$ and $c_1$ in eq.(\ref{eq2-18}) become
\begin{equation}
c_0=1, \quad {\rm and }\quad c_1=4-\delta_1+\frac{\pi^2}{12}-\frac{\gamma^2}{2}
\simeq -2.98961,
\end{equation}
respectively. Here we use  $\delta_1=G^{32}_{23}\left(1|\matrix{
-1 & 0 & 1 \cr
0 & 0 & 0 \cr }\right)
\simeq 1.64549 ..$ from the formulae\cite{wille}
\begin{equation}\label{laplace}
\int_0^{\infty}dv v^{\alpha}e^{-\eta v}G^{31}_{23}\left(\eta v|\matrix{
0 & 1 & \cr
0 & 0 & 0 \cr }\right)=\eta^{-(1+\alpha)}G^{32}_{23}\left(1|\matrix{
-\alpha & 0 & 1  \cr
0 & 0 & 0 \cr }\right).
\end{equation}

Using this distribution function, the mobility $\mu$ defined through
\begin{equation}
\bar v\equiv \int_0^{\infty} dvv\int_0^{2\pi}d\theta v\cos\theta P(v,\theta)
=\mu F_{ex}
\end{equation}
 is given by
\begin{equation}
\mu=\frac{\sqrt{T} }{4\sqrt{3m^3}\zeta}
, 
\end{equation}
where we use $D=\zeta\sqrt{T/3m}$\citep{kawarada}. 

Now, we comment on the dimensionless forms of our result obtained here.
With the aid of (\ref{scale-1}) and (\ref{scale-2}), it is easy to check
$\eta v$ becomes $\sqrt{6}c$ in the scaling form. In
addition , the prefactor $D/g$ in (\ref{eq2-7}) becomes
$D/(g v_0)=1/\sqrt{6\epsilon}$.   
Figure 1 plots the scaled $P_0(c)$ and $P_1(c)$ as functions of
$c=v/v_0$. 
%From Fig.1 both of $P_0$ and $P_1$ have similar tendencies in 
%which $P_n(c)$ with $n=0,1$ becomes negative near $c=0$ and has a
%positive peak around $c=1$. 

\begin{figure}
 \begin{center}
  \includegraphics[width=120mm]{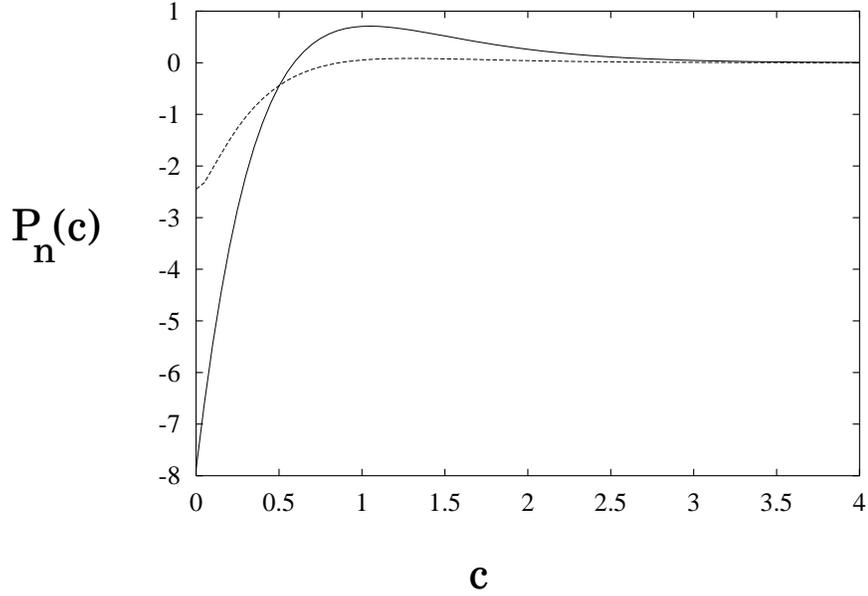}
 \end{center}
\caption{
The scaled $P_0(c)$ (solid line) and $P_1(c)$ (dashed line) as functions 
 of $c=v/v_0$, where $P_n(c)$ denotes $P_0(c)$ or $P_1(c)$ in the figure. 
For $P_1(c)$ we plot ${f^{(1)}}'(c) e^{-\sqrt{6}c}$ to
 remove the effect of the expansion parameter $\epsilon$. }
\end{figure}

\section{Discussion}

Let us discuss our result. First, we can obtain higher order terms
such as $f^{(2)}(v)$ and $f^{(3)}(v)$, because the homogeneous solutions 
in eqs.(\ref{eq2-13a}) and (\ref{eq2-13b}) can be used in any
order. However, these solutions may have complicated forms and we had
better use numerical integrations to represent inhomogeneous
terms. Because of the limitation of the length of this paper, we have
omitted to give the explicit forms of higher order terms here. 

Second, Kawarada and Hayakawa\citep{kawarada} assume that the diffusion
coefficient in the real space is proportional to the diffusion constant
in the velocity space, but this assumption may not be true for our case.  
Let us briefly evaluate the diffusion constant in the real space as
follows. The basic equation is
\begin{equation}
\partial_t P({\bf r},{\bf v},t)=\Gamma P= (\Gamma_0+\Gamma_1)P
\end{equation}
where 
\begin{equation}
\Gamma_0= D\partial_{{\bf v}}(\eta \frac{{\bf v}}{v}+\partial_{{\bf v}}) 
\quad {\rm and } \quad
\Gamma_1= -{\bf v}\cdot \nabla .
\end{equation}
Now, we introduce the projection operator 
\begin{equation}
{\hat P}h({\bf r},{\bf v},t)=\varphi_0(v)\int d{\bf v}h({\bf r},{\bf v},t).
\end{equation}
%with $\varphi_0(v)=\frac{\eta^2}{2\pi}\exp(-\eta v)$.
With the aid of Mori-Zwanzig projection
method\citep{zwanzig,mori}, in general, the distribution function obeys
an identity
\begin{eqnarray}\label{projection}
\partial_t{\hat P}P&=&{\hat P}\Gamma {\hat P}P+{\hat P}\Gamma\int_0^td\tau 
e^{(t-\tau){\hat Q}\Gamma}{\hat Q}\Gamma{\hat P}P(\tau)\nonumber \\
& &+{\hat P}\Gamma 
e^{t{\hat Q}\Gamma}{\hat Q}P(t_0)
\end{eqnarray}
where ${\hat Q}=1-{\hat P}$.
%, and $f_0$ is the initial distributionfunction. 
It is easy to show that the first and the last terms of right
hand side of  (\ref{projection}) are zero if the initial distribution is 
proportional to $\varphi_0(v)$. We also assume that $\Gamma_1$ can be
regarded as the pertubation of $\Gamma_0$ because the distribution
function is unchanged within the mean-free path of particles. Thus, we
may replace $e^{(t-\tau){\hat Q}\Gamma}$ by $e^{(t-\tau){\Gamma}_0}$ in
the second term of right hand side of (\ref{projection}). Noting
\begin{eqnarray}
{\hat Q}\Gamma_1\varphi_0(v)\bar F({\bf r},t)&=&-{\bf v}\cdot\nabla{\varphi}_0(v) \bar F({\bf r},t) \\
\Gamma_0{\bf v}\varphi_0(v)&=&-\zeta \frac{{\bf v}}{v}\varphi_0(v)
\label{eigen}\end{eqnarray}
with $\bar F\equiv {\hat P}P$, eq.(\ref{projection}) may be reduced to
\begin{eqnarray}
\partial_t\bar F&=&-\int d{\bf v}\Gamma \int_0^td\tau e^{(t-\tau)\Gamma_0}
{\bf v}\varphi_0\nabla \bar F \nonumber \\
&\simeq& \nabla \int d{\bf v}\int_0^t d\tau e^{-\frac{\zeta}{v}(t-\tau)}
v^2\varphi_0\nabla \bar F \nonumber \\
&\simeq& \frac{1}{\zeta}\int d{\bf v}v^3\varphi_0(v)\nabla^2\bar F
=D_x \nabla^2 \bar F  
\end{eqnarray}
in the long time limit $t\gg v_0/\zeta$.
Thus, the diffusion coefficient in the real space may be $D_x\propto
D^3/\zeta^4\propto T^{3/2}$. 
We should note that this procedure is not exact because
$t\gg v_0/\zeta$ does not ensure the approximation $\int_0^td\tau
e^{-\zeta t/v}\simeq v/\zeta$ for large $v\gg v_0$. 
However, the treatment may be
plausible and is contradicted with the previous result of simulation of 
granular particles.  

We also note that the behavior of VDF under the influence of $g$ in
granular particles does not coincide with the theoretical prediction
presented here quantitatively. 
As shown in Fig.1, $P_0$ and $P_1$ have  peaks around $c=1$ and 
become negative near $c=0$. The behavior of the
simulation for granular particles is qualitatively similar, but the
peak position is located near $c=0.5$  and
the negativity in the vicinity of $c=0$ is not large
in the simulation.

Thus, we will have to check whether the connection between Langevin
equation with Coulomb friction and the vibrated granular particles
confined in a quasi-two-dimensional box is superficial. 
The detailed quantitative comparisons will be
reported elsewhere.

%We sometimes introduce the effective temperature $T_{eff}$ as the ratio
%$D/\mu$. In our case, the situation is a little strange, because
%$T_{eff}$ becomes
%\begin{equation}
%T_{eff}=\frac{2m^{2}}{\pi T}\frac{1}{(4-3c_0)}
%\propto T^{-1}
%\end{equation}
%Thus, $T_{eff}$ decreases as the granular temperature increases.

%We should note that the relation between the diffusion constant $D$ and
%the temperature $T$ defined by the second moment of the velocity
%distribution function 
%\begin{equation}
%T\equiv \frac{1}{2}m\int d{\bf v}v^2 P({\bf v})=\frac{3m}{ \eta^3}
%+\epsilon T^{(1)}+\cdots
%\end{equation}
%where
%\begin{equation}
%T^{(1)}=
%\frac{1}{4}m \eta^2\int_0^{\infty}dv v^3f^{(1)}(v)e^{-\eta v}
%\end{equation}
%is modified in the existence of the external
%field. With the aid of (\ref{eq2-18}) and (\ref{laplace})  we obtain
%\begin{equation}
%T^{(1)}\eta^{4}=c_0(\delta_2+82-24\gamma-3\gamma^2+\frac{\pi^2}{2}
%-59\ln \eta+3(\ln \eta)^2)+60.
%\end{equation}
%where
%\begin{equation}
%\delta_1=G^{32}_{23}\left(1|\matrix{
%-3 & 0 & 1  \cr
%0 & 0 & 0 \cr }\right).
%\end{equation}

\section{Conclusion}

In conclusion, we have developed the theory of  
Langevin equation with Coulomb friction and obtained the steady
solution of VDF under the influence of
a steady external field. We also discuss the relation between our system 
with the vibrated granular systems
confined in a quasi-two-dimensional box

We would like to thank A. Kawarada for  fruitful discussion.
This work is partially supported by the Grant-in-Aid 
for Scientific Research (Grand No.15540393) of the Ministry of 
Education, Culture, Sports, Science and Technology, Japan.


\begin{thebibliography}{}

% \bibitem[Names(Year)]{label} or \bibitem[Names(Year)Long names]{label}.
% (\harvarditem{Name}{Year}{label} is also supported.)

\bibitem[Kubo92]{Kubo} 
R. Kubo, M. Toda and N. Hashitume; Statistical Physics
	      II. {\it Nonequilibrium Statistical Mechanics}
	      (Springer-Verlag, Berlin, 1992).

\bibitem[Perrson98]{perrson} B. N. J. Perrson, Sliding Friction: {\it Physical
	      Principles and Approximations} (Springer-Verlag, Berlin,
	      1998).

\bibitem[Kawarada04]{kawarada} A. Kawarada and H. Hayakawa, to be published in
	      J. Phys. Soc. Jpn. {\bf 73}, No. 8 (2004) (cond-mat/0404633). 
\bibitem[Arfken95]{arfken} G. B. Arfken and H. J. Weber, Mathematical Methods
		for Physicists, {\it 4th. ed.} (Academic Press,
		SanDiego, 1995).
\bibitem[Meijer41]{meijer} C. S. Meijer, Proc. Nederl. Akad,
		Wetensch. {\bf 44}, 1062 (1941).

\bibitem[Wille86]{wille} L. T. Wille, J. Phys. A {\bf 19}, L313 (1986).

\bibitem[Zwanzig61]{zwanzig} R. W. Zwanzig, 
 Phys. Rev. {\bf 124}, 983 (1961).
\bibitem[Mori65]{mori} H. Mori, Prog. Theor. Phys. {\bf 33}, 423 (1965).

\end{thebibliography}
\end{document}